\documentclass{article}

\usepackage{amsfonts}
\usepackage{amssymb,amsmath}
\usepackage{color}
\usepackage{soul}
\usepackage{graphicx}
\usepackage{caption}
\usepackage{subcaption}

\graphicspath{ {./Figures/} } 

\begin{document}

%%%%%%%%%%%%%%%%%%%%%%%%%%%%%%%%%%%%%%%%%%%%%%%%%%%%%%%

\markboth{A.G. Fletcher, P.J. Murray \& P.K. Maini}{Multiscale modelling of colonic crypt organization and carcinogenesis}

%%%%%%%%%%%%%%%%%%%%%%%%%%%%%%%%%%%%%%%%%%%%%%%%%%%%%%%

\title{MULTISCALE MODELLING OF INTESTINAL CRYPT ORGANIZATION AND CARCINOGENESIS}

\author{ALEXANDER G. FLETCHER\\ Mathematical Institute, University of Oxford, UK
\and PHILIP J. MURRAY\\ Department of Mathematics, University of Dundee, UK
\and PHILIP K. MAINI\\ Mathematical Institute, University of Oxford, UK}

\date{}

%alexander.fletcher@maths.ox.ac.uk
%pmurray@maths.dundee.ac.uk
%philip.maini@maths.ox.ac.uk

\maketitle

%%%%%%%%%%%%%%%%%%%%%%%%%%%%%%%%%%%%%%%%%%%%%%%%%%%%%%%

\begin{abstract}
Colorectal cancers are the third most common type of cancer. 
They originate from intestinal crypts, glands that descend from the intestinal lumen into the underlying connective tissue. Normal crypts are thought to exist in a dynamic equilibrium where the rate of cell production at the base of a crypt is matched by that of loss at the top.
Understanding how genetic alterations accumulate and proceed to disrupt this dynamic equilibrium  is fundamental to understanding the origins of colorectal cancer.
Colorectal cancer emerges from the interaction of biological processes that span several spatial scales, from mutations that cause inappropriate intracellular responses to changes at the cell/tissue level, such as uncontrolled proliferation and altered motility and adhesion. 
Multiscale mathematical modelling can provide insight into the spatiotemporal organisation of such a complex, highly regulated and dynamic system. Moreover, the aforementioned challenges are inherent to the multiscale modelling of biological tissue more generally. 
In this review we describe the mathematical approaches that have been applied to investigate multiscale aspects of crypt behaviour, highlighting a number of model predictions that have since been validated experimentally. 
We also discuss some of the key mathematical and computational challenges associated with the multiscale modelling approach. 
We conclude by discussing recent efforts to derive coarse-grained descriptions of such models, which may offer one way of reducing the computational cost of simulation by leveraging well-established tools of mathematical analysis to address key problems in multiscale modelling. 
\end{abstract}

%%%%%%%%%%%%%%%%%%%%%%%%%%%%%%%%%%%%%%%%%%%%%%%%%%%%%%%


%%%%%%%%%%%%%%%%%%%%%%%%%%%%%%%%%%%%%%%%%%%%%%%%%%%%%%%


%%%%%%%%%%%%%%%%%%%%%%%%%%%%%%%%%%%%%%%%%%%%%%%%%%%%%%%

\section{Introduction}

The past decade has witnessed remarkable progress in experimental studies of intestinal tissue self-renewal and colorectal carcinogenesis. 
This is due in part to increased molecular understanding of intestinal cell biology, as well as advances in techniques such as clonal analysis,\cite{Baker2014Quantification,LopezGarcia2010Intestinal,Snippert2010Intestinal} {\it in vitro} tissue culture\cite{Sato2009Single} and intravital microscopy.\cite{Ritsma2014Intestinal}
Mathematical modelling offers a complementary tool with which to unravel the complex interactions between processes at the intracellular, cellular and tissue scales that underlie the spread and fixation of mutations during the first stages of colorectal cancer. 
Models can be used to develop abstract representations of biological systems, test competing hypotheses, and generate new predictions that can then be validated experimentally. 
To this end, a variety of different individual cell-based and multiscale modelling approaches have recently been developed for studying how processes at the level of a single cell affect the tissue-level behaviour of the intestinal epithelium.
This review aims to highlight the mathematical challenges inherent in multiscale modelling of intestinal self-renewal and colorectal cancer, which serves as a representative example of the more general study of stem cell dynamics, tissue homeostasis and carcinogenesis.

The remainder of the paper is structured as follows. 
In Section~\ref{sec:intro_biology} we summarise the relevant background biology. 
In Section~\ref{sec:intro_modelling} we briefly discuss the variety of mathematical modelling approaches that have been applied to the study of colorectal cancer, placing multiscale models into a wider context. 
In Section~\ref{sec:multiscale_models} we review multiscale modelling efforts, focusing on the competing approaches and on the model predictions that have since been validated experimentally. 
In Section~\ref{sec:challenges} we highlight some of the key mathematical and computational challenges associated with the multiscale modelling approach and discuss recent efforts to derive coarse-grained descriptions of such models, which aim to improve analytic tractability while retaining the essential features of the underlying models. 
Finally, in Section~\ref{sec:conclusions} we offer concluding remarks on future directions for this work.

%%%%%%%%%%%%%%%%%%%%%%%%%%%%%%%%%%%%%%%%%%%%%%%%%%%%%%%

\section{Background biology} \label{sec:intro_biology}

\begin{figure}
\centering
\begin{subfigure}[b]{0.4\textwidth}
\centering
\includegraphics[scale=0.2]{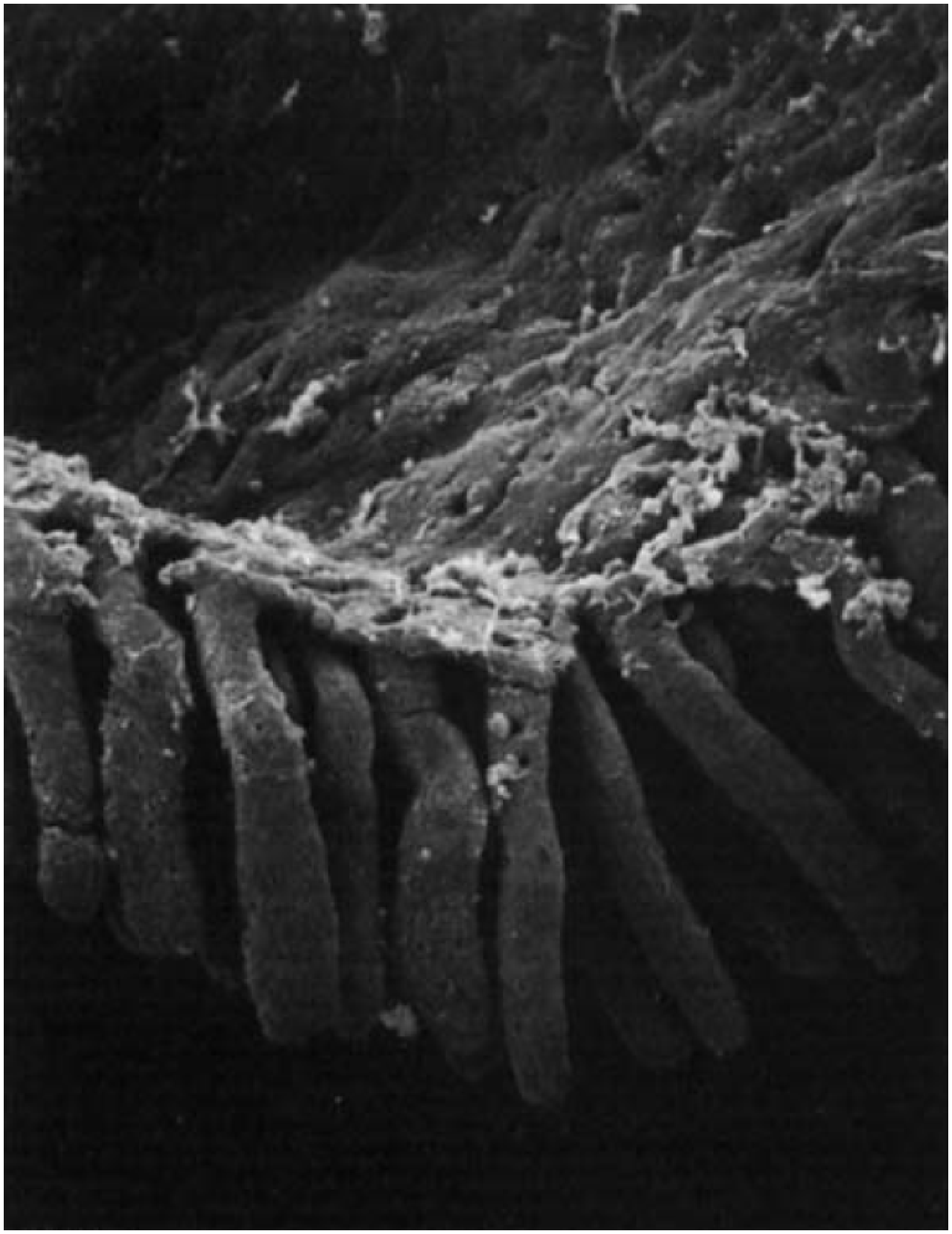}
\caption{}
\label{fig:CryptSchematicFigureA}
\end{subfigure}%
~
\begin{subfigure}[b]{0.6\textwidth}
\centering
\includegraphics[scale=0.4]{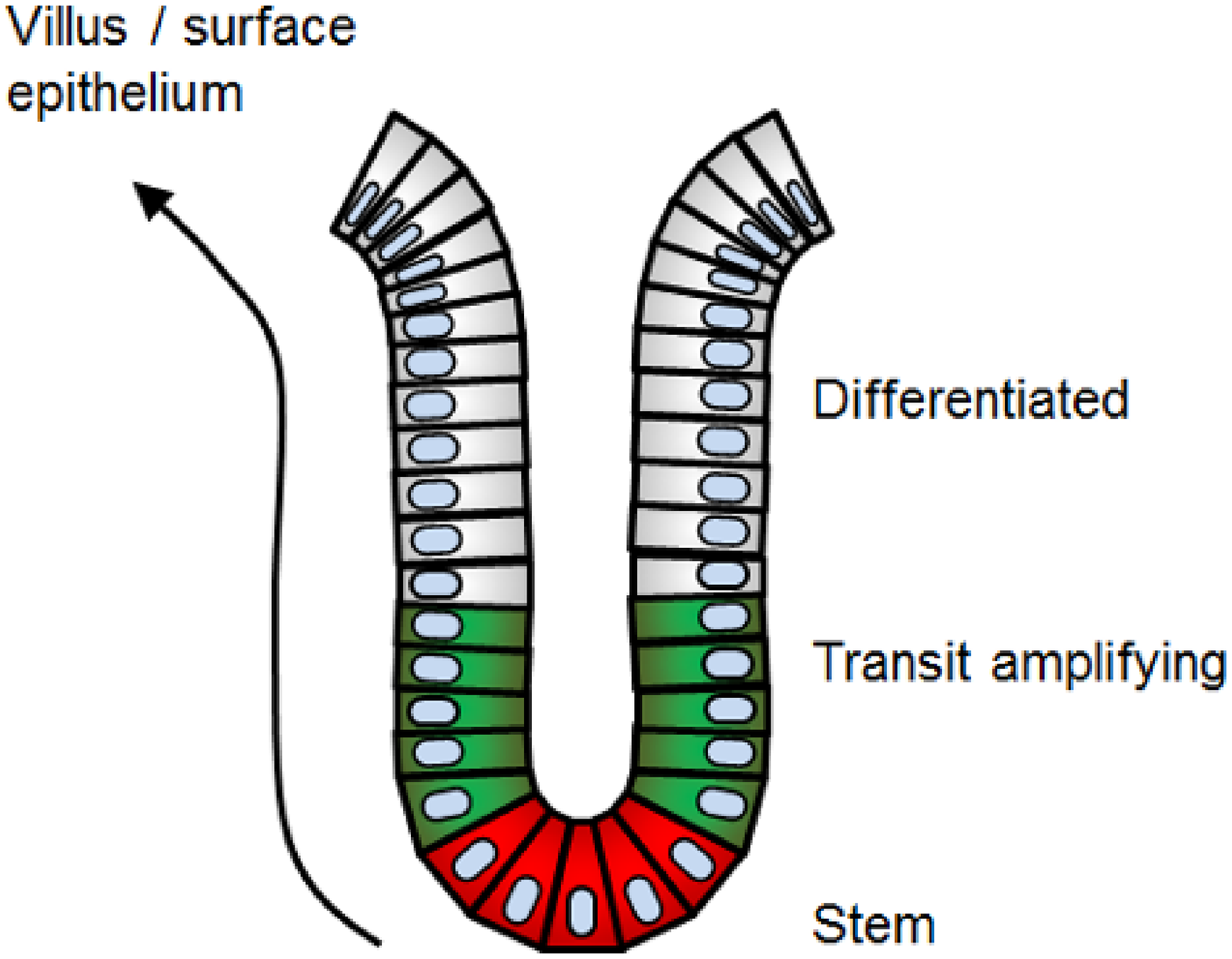}
\caption{}
\label{fig:CryptSchematicFigureB}
\end{subfigure}
~
\begin{subfigure}[b]{\textwidth}
\centering
\includegraphics[scale=0.25]{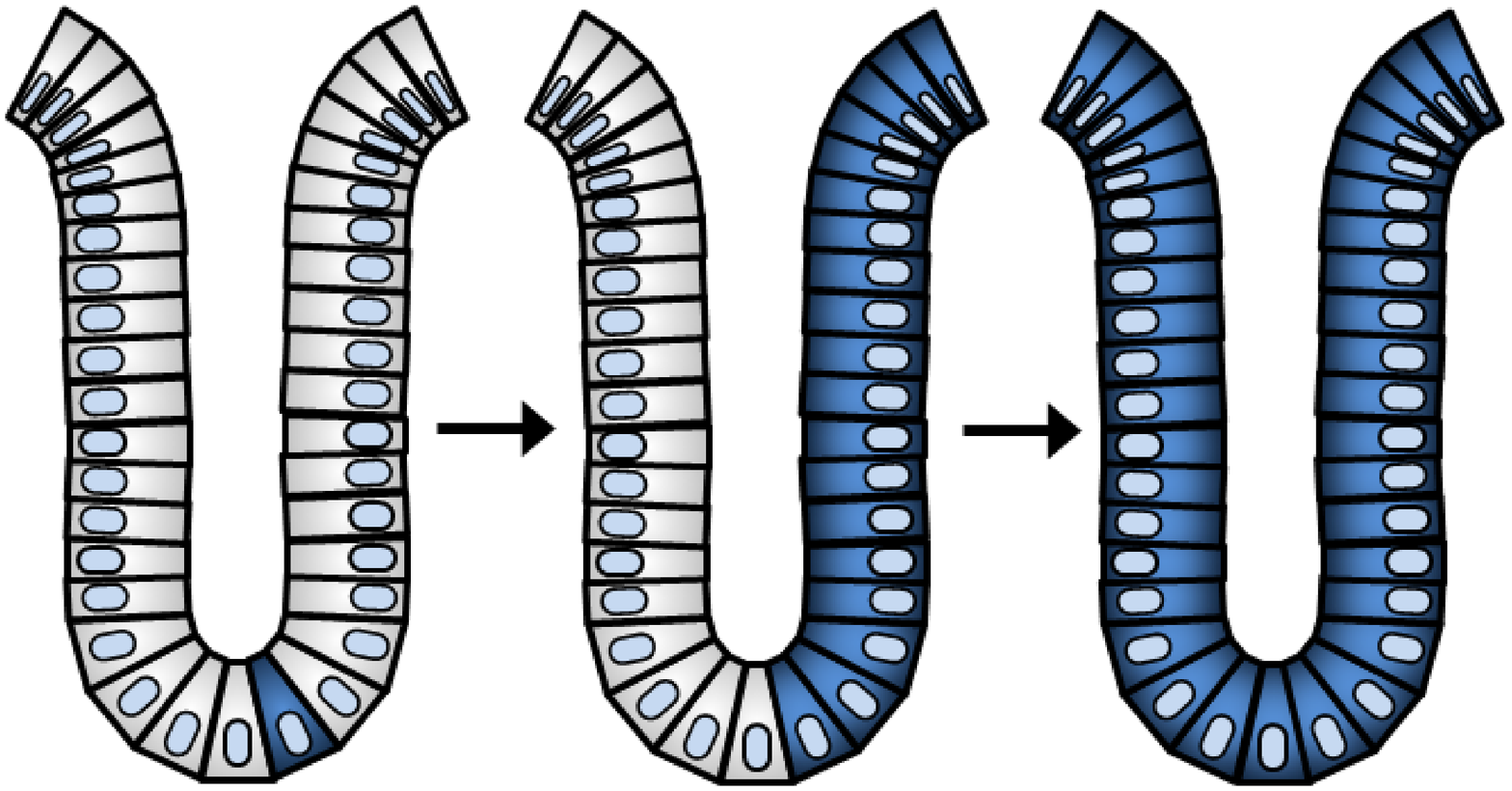}
\caption{}
\label{fig:CryptSchematicFigureC}
\end{subfigure}
~
\begin{subfigure}[b]{\textwidth}
\centering
\includegraphics[scale=0.25]{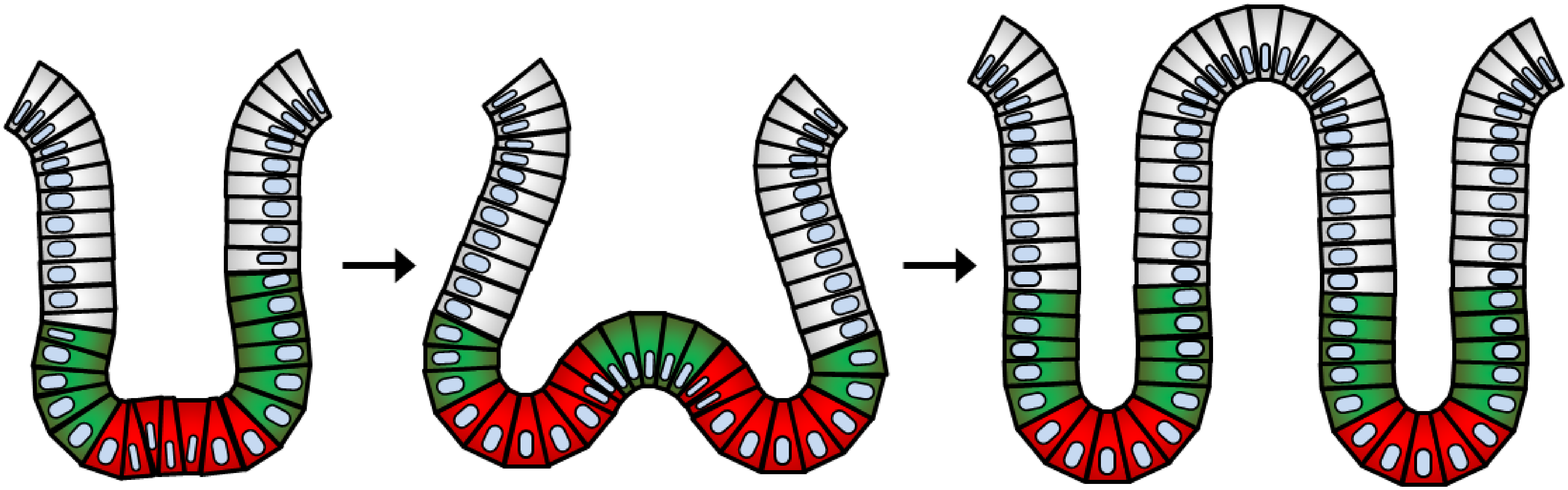}
\caption{}
\label{fig:CryptSchematicFigureD}
\end{subfigure}
\vspace*{8pt}
\caption{(a) Scanning electron micrograph of murine colonic epithelium, showing the structural crypt organization. 
(b) Schematic of the spatial organization of the colonic crypt. Stems cells (red) residing at the crypt base generate rapidly proliferating transit amplifying cells (green), which ascend the crypt and differentiate into the mature cell lineages (white) occupying the surface epithelium (in the colon) or adjacent villi (in the small intestine). 
(c) Schematic of the process of monoclonal conversion, where the progeny of a single stem cell (blue) replaces all other cells in the crypt as a result of neutral drift. 
(d) Schematic of the process of crypt fission, where excess proliferation at the crypt base results in buckling, basal bifurcation and longitudinal division; this process facilitates the spread of mutations beyond individual crypts. 
Image (a) reproduced with permission from Magney et al.\protect\cite{Magney1986Scanning} \copyright Wiley (1986). 
}
\label{fig:CryptSchematicFigure}
\end{figure}

Colorectal cancers initiate in the epithelial tissue that covers the luminal surface of the intestinal tract. 
The rapid renewal of this epithelium involves a coordinated programme of cell proliferation, migration and differentiation, which begins in the tiny intestinal crypts, finger-like invaginations of the intestinal mucosa lined by a monolayer of epithelial cells (Fig.~\ref{fig:CryptSchematicFigureA}). 
Crypts function to absorb water and exchange electrolytes from the faeces, and to produce mucus to lubricate faeces as they move through the large intestine (colon).
Colorectal carcinogenesis occurs as a consequence of changes that disrupt normal crypt dynamics. 
Thus, the tissue architecture and dynamics of intestinal crypts are central to the origins of colorectal cancer.\cite{Humphries2008Colonic}

A population of stem cells located at the base of each crypt divide frequently to generate transit amplifying cells that migrate up the crypt wall,\cite{Barker2007Identification} as shown in Fig.~\ref{fig:CryptSchematicFigureB}. 
As they ascend, these cells undergo a sequence of rapid divisions before they stop
dividing and differentiate terminally into  specialised progeny, such as colonocytes, enteroendocrine cells or Goblet cells, that occupy the upper portion of the crypt.\cite{Brittan2004Stem,Sangiorgi2008Bmi1} 
In the colon, cells at the crypt orifice either undergo apoptosis or
are shed into the lumen and transported away, while in the small intestine they move onto an adjacent villus structure. 
Crypts may be generated by a process called crypt fission, in which a crypt bifurcates basally and then divides longitudinally. 
Crypt fission occurs sporadically in the adult intestine as well as in response to epithelial damage.\cite{Totafurno1987Crypt}
Crypt extinction is also possible and occurs if all actual and potential stem cells are eliminated, for example after
exposure to radiation or cytotoxic drugs.\cite{Cheng1986Crypt}

Intestinal stem cells are in continual neutral competition with one another to retain a privileged position within the niche at the crypt base.\cite{Barker2014Adult} 
Each stem cell division results, on average, in the loss and replacement of an individual stem cell lineage.\cite{Baker2014Quantification,Kozar2013Continuous,LopezGarcia2010Intestinal,Snippert2010Intestinal} 
A consequence of this competition is that over time, the progeny of a single stem cell within a crypt can take over the entire crypt. 
This process is termed monoclonal conversion, since the resulting crypt consists of a single clonal population. 
Mutant monoclonal crypts are likely to constitute the earliest stage of colorectal adenomas, which then expand further through crypt fission rather than unconstrained growth of cells {\it per se}.\cite{Greaves2006Mitochondrial,Preston2003Bottom,Thirlwell2010Clonality,Wong2002Histogenesis}
The rate of fission is significantly altered in such mutant crypts; for example, the basal rate of crypt fission in the mouse colon is increase 30-fold by the oncogenic KRAS mutation.\cite{Snippert2014Biased} 
This is likely due to the increased stress on crypt walls due to the persistence and excess proliferation of mutant cells, which can cause crypt buckling and fission.

A number of biochemical signalling pathways have been implicated in the regulation of intestinal stem cell and crypt dynamics. Of particular importance in the context of colorectal cancer is Wnt  signalling, which is initiated when extracellular Wnt factors bind to
specific receptors on the cell surface. 
This triggers a cascade of intracellular reactions that leads, via the transcription of Wnt target genes, to the regulation of proteins involved
in cell-cycle control, migration and apoptosis. 
Mutations in key components of the Wnt
pathway, such as APC and $\beta$-catenin, have been shown to be the first step in colorectal carcinogenesis in the majority of colorectal cancers, including most hereditary cases.\cite{Ilyas2005Wnt} 
Notably, the stem cell niche in the colonic crypt is regulated by a spatial gradient of extracellular Wnt factors along the crypt axis.\cite{Hirata2013Dose,vanDeWetering2002Beta}

Despite the central role of stem cells in maintaining
the integrity of the intestinal epithelium, for many years their precise number and location within the crypt has remained unclear, due to a lack of unique marker genes and the absence of stem-cell assays. 
As the transgenic approach used successfully in mice\cite{Barker2007Identification,McDonald2006Clonal,Winton1988Clonal} is impractical for human studies, it has remained uncertain if and how the biology of the human intestinal crypt is mirrored by its murine counterpart. 
Further, the mutagenic manipulations required to observe crypt heterogeneity may alter the normal behaviour of stem cells. 

As the above observations demonstrate, the intestinal crypt is a complex, highly regulated system whose dynamics under normal and pathological conditions are difficult to interpret using verbal arguments alone.
Mathematical modelling offers a powerful tool that can provide mechanistic insights that complement and reinforce knowledge acquired from molecular, histological and live-imaging studies. 
Models can be used to integrate our assumptions and measurements within a consistent theoretical framework, test competing hypotheses {\it in silico}, and generate new predictions that can then be validated experimentally.
In the next section we provide a brief overview of mathematical approaches that have been used to describe crypt dynamics, before focussing in Section~\ref{sec:multiscale_models} on multiscale models of crypt organization and carcinogenesis.

%%%%%%%%%%%%%%%%%%%%%%%%%%%%%%%%%%%%%%%%%%%%%%%%%%%%%%%

\section{Modelling approaches for colorectal cancer} \label{sec:intro_modelling}

Mathematical modelling has been used to investigate aspects of colorectal cancer for over half a century. 
The first such model, by Armitage and Doll,\cite{Armitage1954Age} was used to explain experimental data indicating a power law relationship between age and cancer incidence. 
This model highlights the multistage nature of colorectal cancer, by predicting that approximately six successive mutations in a single cell are required for carcinogenesis to occur. 

As the quality of cellular-resolution experimental data has improved over the last two decades, a variety of mathematical models have been used to study cell population dynamics in the colonic crypt. 
These include both deterministic\cite{Boman2001Computer,Johnston2007Mathematical} and stochastic\cite{Loeffler1997Clonality,Nowak2003Linear} models. 
As an illustrative example we highlight the work of Johnston et al.\cite{Johnston2007Mathematical}, who analyse and compare age-structured and continuous models of crypt population dynamics and propose two feedback mechanisms that may regulate cell numbers to maintain crypt homeostasis. 
Here we focus on their continuous model, which considers timescales much greater than the typical crypt cell cycle time and populations large enough to warrant a continuous approach. 
Letting $N_{0}(t)$, $N_{1}(t)$ and $N_{2}(t)$ denote the size of the stem, this model comprises the set of coupled ordinary differential equations
\begin{align}
\label{eq:johnston_0} \frac{\mathrm{d}N_{0}}{\mathrm{d}t} & = (\alpha_{3} - \alpha_{1} - \alpha_{2})N_{0}, \\
\label{eq:johnston_1} \frac{\mathrm{d}N_{1}}{\mathrm{d}t} & = (\beta_{3} - \beta_{1} - \beta_{2})N_{1} + \alpha_{2}N_{0}, \\
\label{eq:johnston_2} \frac{\mathrm{d}N_{2}}{\mathrm{d}t} & = \beta_{2} N_{1} - \gamma N_{2}.
\end{align}
Here the parameters $\alpha_{1}$, $\alpha_{2}$ and $\alpha_{3}$ (respectively $\beta_{1}$, $\beta_{2}$ and $\beta_{3}$) denote the per-capita rates of stem (respectively transit amplifying) cell death, differentiation and proliferation, while $\gamma$ denotes the per-capita rate of removal of differentiated cells due to sloughing. 
If these rates are assumed to take constant values, then given an initial crypt population $N_{0}(0) = \hat{n}_{0}$, $N_{1}(0) = \hat{n}_{1}$, $N_{2}(0) = \hat{n}_{2}$, the system of equations~\eqref{eq:johnston_0}--\eqref{eq:johnston_2} has the straightforward solution
\begin{align}
N_{0}(t) & = \hat{n}_{0} e^{\alpha t}, \\
N_{1}(t) & = A e^{\alpha t} + (\hat{n}_{1} - A) e^{\beta t}, \\
N_{2}(t) & = Be^{\alpha t} + Ce^{\beta t} + (\hat{n}_{2} - B - C)e^{-\gamma t},
\end{align}
where $\alpha = \alpha_{3} - \alpha_{1} - \alpha_{2}$, $\beta = \beta_{3} - \beta_{1} - \beta_{2}$ and
\begin{align}
A = \frac{\alpha_{2} \hat{n}_{0}}{\alpha - \beta}, \quad B = \frac{\beta_{2}A}{\gamma + \alpha}, \quad \mbox{and} \quad C = \frac{\beta_{2}(\hat{n}_{1} - A)}{\gamma + \beta}.
\end{align}
This solution exhibits exponential growth or decay according to the values of $\alpha$, $\beta$ and $\gamma$. 
Johnston et al.\cite{Johnston2007Mathematical} note that such structural instability is unrealistic, since a crypt must be able to maintain homeostasis in the face of fluctuations in rates of proliferation and differentiation, for example due to stochasticity. 
Some form of feedback is required to ensure such robustness. 
The authors examine two possibilities in their model, feedback on proportions of cells differentiating and on rates of cell division, and deduce that the latter form offers a more feasible explanation for observed lag phases in crypt dynamics after mutations occur, and thus the existence of benign tumours prior to the onset of carcinogenesis. 
Such work highlights the utility of simple, analytically tractable population dynamics models in this field.

While earlier models focus on temporal dynamics only, increasingly modellers have sought also to understand the spatial organisation of the crypt. 
The classical approach to modelling spatial dynamics is to treat the tissue as a continuum.\cite{Figueiredo2011Convection}
As such models average over length scales that are much larger than the typical diameter of a cell, processes such as proliferation and adhesion, which may become disrupted in cancerous cells, can be considered only in an averaged sense. 
Such models do not allow the straightforward incorporation of subcellular processes, such as signalling pathways or protein-level descriptions of cell-cell adhesion. 
When studying colonic crypt organisation and carcinogenesis, a further challenge to the use of classical, continuum models is the relatively small numbers of cells in the tissue of interest; intestinal crypts contain around 300 cells in mice and 2000 cells in humans.

As a result of these limitations, over the last few years a number of authors have adopted a discrete approach to modelling crypt cell dynamics in health and disease. 
Several discrete approaches have been applied to model crypt cell population dynamics and combine descriptions of intracellular and cellular processes. 
These vary in complexity from fixed-lattice cellular automata and the cellular Potts model to off-lattice cell-centre and vertex models. 
Such multiscale models allow for the integration of disparate experimental data and the prediction of system-level behaviour, but are more complicated and less amenable to mathematical analysis than their single-scale counterparts.\cite{Fletcher2011Multiscale}

Before proceeding to focus on multiscale modelling, we note the availability of recent review articles on other aspects of intestinal tissue renewal and colorectal cancer. 
These include reviews focused on the biological insights gained by compartment models of crypt dynamics,\cite{Carulli2014Unraveling} the contrasting approaches and software implementations for spatial computational models,\cite{DiMatteis2013Review} and the burgeoning links between modelling and experiments in this field.\cite{Kershaw2013Colorectal}

%%%%%%%%%%%%%%%%%%%%%%%%%%%%%%%%%%%%%%%%%%%%%%%%%%%%%%%

\section{Multiscale models of crypt organization and carcinogenesis} \label{sec:multiscale_models}

We next focus on a small number of representative individual cell-based and multiscale models of crypt dynamics, broadly in order of mathematical complexity.

%%%%%%%%%%%%%%%%%%%%%%%%%%%%%%%%%%%%%%%%%%%%%%%%%%%%%%%

\subsection{Stem cell dynamics in the crypt} \label{sec:cellular_automata}

One of the first theoretical crypt studies to explicitly model the spatiotemporal dynamics of individual cells is by Loeffler and coworkers,\cite{Loeffler1986Intestinal} who use a stochastic cellular automaton approach. 
In cellular automata, physical space is discretized using a fixed lattice on which cells can occupy individual sites. 
The rules for cell movement are formulated in terms of cells moving between lattice sites. 
The state of each cell in the system is updated over discrete time steps using automaton rules and the state of the cell's neighbourhood at the previous time step.\cite{Block2007Classifying,Lee1995Cellular}
It is possible to efficiently simulate the interactions of large numbers of cells in cellular automata; this advantage comes at the expense of an artificial spatial lattice anisotropy and hence discontinuous cell motion. 

In their model, Loeffler and colleagues\cite{Loeffler1986Intestinal,Paulus1993Differentiation} demonstrate that the spatial arrangement of proliferating columnar cells, as well as mucus-producing Goblet cells, can be explained by the concept of a cellular pedigree. 
This assumes that a self-maintaining stem cell generates transit amplifying cells, which undergo a well-defined series of cell divisions before ceasing proliferation. 
Under this hypothesis, regulatory control is required only in the stem cell region. 
Loeffler et al.\cite{Loeffler1986Intestinal} also investigate how the probability of a stem cell generating zero, one or two stem cells upon division affects the crypt structure. 
The authors initially assume these probabilities to take constant values, thus neglecting possible feedback mechanisms, and find that the probability of asymmetric division should lie in the range (0.8, 0.95), assuming a 24 h cell cycle time. 
This prediction proved to be inconsistent with more recent data on the time scale over which mutated crypts become monoclonal,\cite{Winton1988Clonal} 
highlighting how such models may be validated as new measurements become available.

In further work, Loeffler and colleagues seek to modify their crypt model so as to recapitulate the observed timescales of monoclonal conversion.\cite{Paulus1993Differentiation} 
The authors incorporate a state dependence into the probabilities of stem cell division and stem cell cycle time, in which stem cell growth is inhibited if a crypt already contains numerous stem cells. 
Crypt fission is assumed to occur at some threshold number of stem cells per crypt. 
Their modified model is shown to be consistent with data on crypt fission and monoclonal conversion times. 
This research demonstrates that some form of autoregulatory feedback is likely to modulate rates and modes of stem cell division in response to fluctuating environmental conditions, reflected for example by varying crypt cell numbers. 

%%%%%%%%%%%%%%%%%%%%%%%%%%%%%%%%%%%%%%%%%%%%%%%%%%%%%%%

\subsection{Crypt homeostasis and monoclonal conversion} \label{sec:off_lattice_models}

The first off-lattice model of crypt dynamics was developed by Meineke et al.\cite{Meineke2001Cell} 
We choose to focus on the mathematical details of this model as it forms the basis of several more recent crypt models.\cite{vanLeeuwen2009Integrative,Mirams2012Theoretical,Fletcher2012Mathematical,Buske2011Comprehensive,Dunn2012TwoDimensional} 
Unlike lattice-based cellular automata, off-lattice models allow continuous cell positions and enable the study of mechanical effects on cell populations in a straightforward manner. 
Off-lattice models offer a more physically realistic view of cell population dynamics, and may be parameterized by biophysical and kinetic parameters that can be determined experimentally. 

Meineke et al.\cite{Meineke2001Cell} model the crypt as a cylinder, which is unrolled to give a flat rectangular surface with periodic boundary conditions on the left and right sides. 
There are two basic components to the authors' description of cell mechanical interactions. 
The first step is to decide which cells are neighbours of each other, the second is to determine the forces transmitted by these neighbours. Each cell $i$ is represented by a point $\mathbf{x}_i \in \mathbb{R}^{2}$, which may be interpreted as the centre of the nucleus. 
Meineke et al.\cite{Meineke2001Cell} assume that cell movement within the crypt is driven by mitotic activity. 
New-born cells force nearby cells to move away, causing a pressure-driven passive movement primarily up the crypt axis. 
The authors model the local repelling and attracting intercellular forces by a network of springs connecting neighbouring cells. 
The total force $\mathbf{F}_i (t)$ acting on a cell $i$ at time $t$ is equal to the sum of all forces coming from the springs of all neighbouring cells $j \in \mathcal{N}_i(t) $ adjacent to $i$ at that time,
\begin{align}
\label{eom_springy} \mathbf{F}_i (t) = \mu \sum_{j \in \mathcal{N}_i(t)}\mathbf{\hat{x}}_{ij}(t) \left( ||\mathbf{x}_{ij}(t)|| - s_{ij}(t) \right),
\end{align}
where $\mu$ is the spring constant, $ \mathbf{x}_{ij}(t)=\mathbf{x}_i(t)-\mathbf{x}_j(t)$ is the vector from $i$ to $j$ at time $t$, $\mathbf{\hat{x}}_{ij}(t)$ is the corresponding unit vector and $s_{ij}$ is the natural separation (spring length) between cells $i$ and $j$. 
For simplicity, all cells are assumed to have identical mechanical properties. 

Adopting the simplifying assumption that inertial terms are small compared to dissipative terms, Meineke et al.\cite{Meineke2001Cell} arrive at first-order dynamics with the evolution of each cell determined by
\begin{align}
\label{eq:eom_meineke} \frac{\mathrm{d}\mathbf{x}_i}{\mathrm{d}t} & = \frac{1}{\eta_i}\mathbf{F}_{i}(t),
\end{align}
where $\eta$ is the damping constant. 
Equation~\eqref{eq:eom_meineke} is solved numerically by using a simple forward Euler discretization, so that the effective displacement within a small time interval $\Delta t$ is given by
\begin{align}
\mathbf{x}_i(t+\Delta t) = \mathbf{x}_i(t) + \frac{1}{\eta_i}\mathbf{F}_i(t)\Delta t.
\end{align}
As this numerical method is explicit, a sufficiently small time step must be chosen. 

After any cell movement, the cell connectivity network is updated using a Delaunay triangulation and  cell boundaries are determined by a Voronoi tessellation, which generates polygonal cell packing. 
This approach is motivated by experimental evidence that columnar cell boundaries have a convex polygonal shape of about equal area, as seen in projection from the intestinal lumen.
In the model, stem cells are restricted to lie at the bottom of the crypt, and asymmetric stem cell division is assumed. 

Meineke et al.\cite{Meineke2001Cell} implement cell division in their model as follows. 
Upon division, each daughter cell is assigned a random cell cycle time drawn from a suitable distribution. 
When a cell is ready to divide, a new cell is introduced a short distance $\varepsilon$ away from the parent cell, at a random angle. 
To account for cell growth, and to avoid an unrealistically large force between these two neighbouring cells, the rest length of the spring between them is chosen to increase linearly from $\varepsilon$ to its normal value $\ell$ over the first hour after division,
\begin{align}
s_{ij}(t) & = \left\{ \begin{array}{ll}
\varepsilon + (\ell - \varepsilon)(t - t_d), & t \leq t_d + 1, \\
\ell, & t > t_d + 1,
\end{array}
\right.
\end{align}
where $t_d$ denotes the time at which the parent cell divides.

Considering cell proliferation, Meineke et al.\cite{Meineke2001Cell} follow earlier work\cite{Loeffler1986Intestinal} by employing the cellular pedigree concept. 
Their off-lattice model reproduces data from crypt labelling experiments as well as the cellular automaton model of Loeffler et al.\cite{Loeffler1986Intestinal,Paulus1993Differentiation} without the need for the state-dependent cell cycle assumption. 
Whilst resolving a number of issues with previous modelling frameworks, the model by Meineke et al.\cite{Meineke2001Cell} is unable to reproduce the broad, wavy ribbons of clonal populations observed experimentally.\cite{Taylor2003Mitochondrial} 
This model limitation is a consequence of the assumption that stem cell positions are fixed, which enforces asymmetric division and thereby prevents propagation of a labelling mutation within the stem-cell niche. 

To address this and other issues, van Leeuwen et al.\cite{vanLeeuwen2009Integrative} extend the model by considering the role of Wnt signalling in the crypt. 
The inclusion of a Wnt gradient along the crypt axis that defines the stem cell niche allowed the assumption that stem cell positions are fixed to be relaxed. 
The local Wnt concentration is assumed to determine intracellular gene expression levels affecting the cell cycle, with high levels of Wnt at the crypt base driving cell proliferation and defining the stem cell niche. 
A threshold of Wnt signalling is assumed to be necessary for mitosis to occur. 
This crypt model represents the first attempt to link phenomena at the subcellular, cellular, and tissue levels of organization. 
In the model, Wnt signalling is assumed to control cell cyle progression, as well as mechanical interactions through the incorporation of contact edge-dependent cell-cell adhesion and cell size-dependent cell-matrix adhesion.
The model recapitulates the observed spatial variation in nuclear $\beta$-catenin levels along the crypt axis. 
Notably, the model exhibits a gradual change in cell cycle times along the crypt axis, which suggests that, in contrast to the simple classification of `stem' and `non-stem', cells may in reality adopt a more continuous representation of differentiation. 

\begin{figure}
\centering
\includegraphics[width=\textwidth]{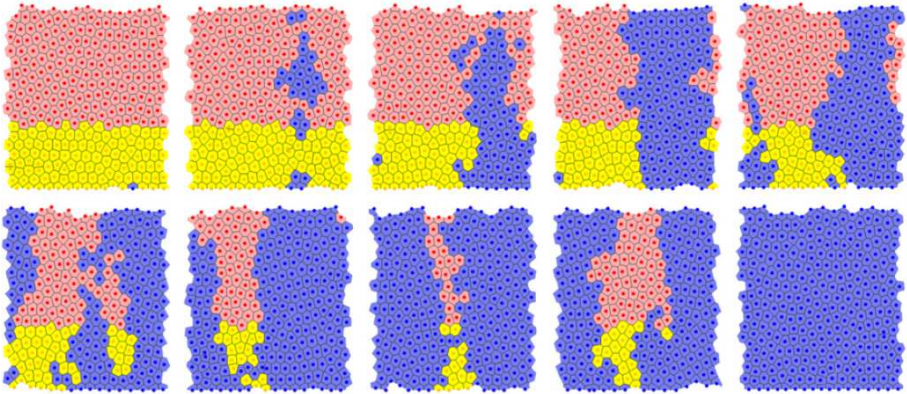}
\caption{
Evolution of a mutant in the crypt model by Mirams et al.\protect\cite{Mirams2012Theoretical}, based on earlier work by Meineke et al.\protect\cite{Meineke2001Cell} and van Leeuwen et al.\protect\cite{vanLeeuwen2009Integrative}. 
A cell with a neutral mutation is introduced near the base of a homeostatic crypt and its progeny are tracked over time. 
Simulation snapshots are shown going left to right then top to bottom. 
Non-mutant cells are coloured yellow or red according to whether they are proliferating or differentiated, while all mutant cells are coloured blue. 
Image adapted with permission from Mirams et al.\protect\cite{Mirams2012Theoretical} \copyright Elsevier.
}
\label{fig:MiramsCryptModelFigure}
\end{figure}

By relaxing the assumption that stem cell positions are fixed, the van Leeuwen et al.\cite{vanLeeuwen2009Integrative} model  recapitulates the observed processes of clonal expansion and monoclonal conversion (Fig.~\ref{fig:MiramsCryptModelFigure}). 
This aspect of crypt behaviour was studied in more detail by Mirams et al.\cite{Mirams2012Theoretical}, who show theoretically that the probability of crypt domination by a mutant population is increased if Wnt signalling is raised near the base of the crypt. 
This emergent property incorporating wild type and mutant cells, different levels of Wnt signalling, proliferation, and adhesion offers a possible explanation for the experimental observation that basally located stem cells are preferential oncogenic targets.\cite{Barker2009Crypt}

\begin{figure}
\centering
\includegraphics[width=0.7\textwidth]{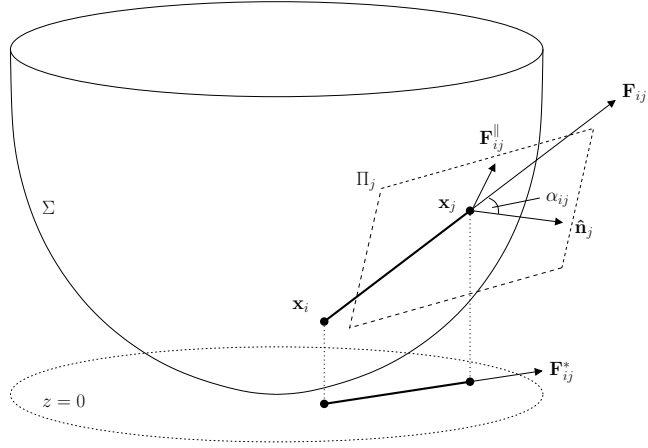}
\vspace*{8pt}
\caption{Schematic illustrating the force calculation in a plane projection model of the crypt developed by Fletcher et al.\protect\cite{Fletcher2012Mathematical}. See main text for details.
}
\label{fig:FletcherCryptModelFigure}
\end{figure}

In the model by van Leeuwen et al.\cite{vanLeeuwen2009Integrative}, the assumption of a cylindrical geometry can lead to unrealistic cell dynamics, especially near the bottom of the crypt. 
In more recent work, Fletcher et al.\cite{Fletcher2012Mathematical} guarantee  a small number of stem cells at the bottom of the crypt by utilising a more realistic crypt geometry. 
This is achieved by projecting the crypt surface, approximated by a surface of revolution
\begin{align}
\Sigma = \{ (r, \theta, z) \in \mathbb{R}^3 \; : \; z = f(r) \},
\end{align}
onto the plane $z=0$. 
This allows the planar model to be used still, facilitating efficient simulation, but with a different metric, determined by the projection.
In more detail, let $\mathbf{F}_{ij}$ define the spring force between two cells $i$ and $j$ given in the summand in equation~\eqref{eom_springy}, and let this force act in three-dimensional space. 
Further, let $\Pi_{j}$ and $\mathbf{n}_{j}$ respectively denote the tangent plane and outward unit normal to $\Sigma$ at the point $\mathbf{x}_{j}$.
Constraining cells to lie on $\Sigma$ results in a reaction force $\mathbf{R}_j$ acting in the direction $-\mathbf{\hat{n}}_j$, whose magnitude balances the component of the force $\mathbf{F}_{ij}$ normal to $\Pi_j$, denoted $\mathbf{F}_{ij}^{\perp}$. 
The component of the force $\mathbf{F}_{ij}$ parallel to $\Pi_j$, denoted $\mathbf{F}_{ij}^{\parallel}$, can be found by taking cross products. 
Using Lagrange's formula for triple vector products, and letting $\mathbf{\hat{e}}_{z}$ denote the unit vector in the $z$ direction, Fletcher et al.\cite{Fletcher2012Mathematical} arrive at the following expression for the projection of $\mathbf{F}_{ij}$ onto the plane $z=0$:
\begin{align}
 \mathbf{F}_{ij}^* & = \mathbf{F}_{ij} - (\mathbf{F}_{ij} \cdot \mathbf{\hat{n}}_j) \mathbf{\hat{n}}_j - (\mathbf{F}_{ij} \cdot \mathbf{\hat{e}}_z) \mathbf{\hat{e}}_z + (\mathbf{F}_{ij} \cdot \mathbf{\hat{n}}_j) (\mathbf{\hat{n}}_j \cdot \mathbf{\hat{e}}_z) \mathbf{\hat{e}}_z.
\end{align}
A diagram illustrating this calculation is shown in Fig.~\ref{fig:FletcherCryptModelFigure}.

The geometrical approximation adopted by Fletcher et al.\cite{Fletcher2012Mathematical} is valid as long as the points are sufficiently close, which is likely the case for packed cells, and introduces a small error which increases with height up the crypt. 
This is in contrast to the cylindrical crypt model,\cite{Meineke2001Cell,vanLeeuwen2009Integrative} in which the error in approximating the shape of the crypt is greatest at the bottom, the region which is critical to the process of monoclonal conversion.
Using this model, Fletcher et al.\cite{Fletcher2012Mathematical} investigate different hypothesised modes of stem cell division by comparing the typical time scale for monoclonal conversion under each hypothesis.
The authors find that the probability and timescale for
this process to occur are sensitive to the precise
geometry of the stem cell niche and to the Wnt stimulus threshold necessary for mitosis to occur. 

%%%%%%%%%%%%%%%%%%%%%%%%%%%%%%%%%%%%%%%%%%%%%%%%%%%%%%%

\subsection{Cell fate determination and lineage specification}

The previous examples highlight how multiscale modelling has yielded mechanistic insight into crypt stem cell dynamics and monoclonal conversion, and how these processes may be influenced by intracellular processes relating to the Wnt pathway. 
However, cell proliferation and differentiation in the crypt is not dependent solely on the presence of a Wnt ligand gradient. 
Other signals, such as Notch, may act in combination with Wnt signalling in this context.\cite{Fre2005Notch,vanEs2005Notch} 
Recent crypt models have sought to understand how the interplay of these signals  regulates cell proliferation.
In particular, following an approach similar to that of van Leeuwen et al.\cite{vanLeeuwen2009Integrative}, Buske et al.\cite{Buske2011Comprehensive} study the combined influence of Wnt and Notch signalling on cell proliferation and differentiation within the first three-dimensional cell-based model of the crypt. 
Notably, the authors also propose the first explicit representation of the basement membrane as a provider of mechanical integrity for the crypt, describing this by a three-dimensional triangular mesh of elastic fibres. 
In contrast to the Delaunay description of cell neighbours employed by Meineke et al.\cite{Meineke2001Cell}, the authors use an `overlapping spheres' approach. 
Here, the radius $R_{i}$ of each cell is accounted for in addition to its position $\mathbf{x}_{i}$. 
By considering changes to cell adhesion, deformation and compression energy terms arising from cell movements and interactions, the authors arrive at a set of coupled ordinary differential equations of the form
\begin{align}
\eta_{BM} \frac{\mathrm{d}\mathbf{x}_{i}}{\mathrm{d}t} + \sum_{j \in \mathcal{N}_{i}(t)} \eta_{C} A_{i,j}^{C} \left( \frac{\mathrm{d}\mathbf{x}_{i}}{\mathrm{d}t} - \frac{\mathrm{d}\mathbf{x}_{j}}{\mathrm{d}t} \right) & = \mathbf{F}_{i}, \\
\eta_{VO} \frac{\mathrm{d}R_{i}}{\mathrm{d}t} + \sum_{j \in \mathcal{N}_{i}(t)} \eta_{C} A_{i,j}^{C} \left( \frac{\mathrm{d}R_{i}}{\mathrm{d}t} + \frac{\mathrm{d}R_{j}}{\mathrm{d}t} \right) & = G_{i},
\end{align}
where the coefficients $\eta_{BM}$, $\eta_{VO}$, $\eta_{C}$ describe friction between a cell and the basement membrane, in response to volumetric changes and between two cells, respectively. 
Here $A_{i,j}^{C}$ denotes the contact area between two overlapping spherical cells $i$, $j$, while $\mathbf{F}_{i}$ and ${G}_{i}$ denote generalized forces acting on cell $i$. 
It is interesting to compare this more detailed biophysical description of cell movement with the simpler spring-based approach discussed above; we return to the issue of model comparison in Section~\ref{sec:comparing_models}.

\begin{figure}
\centering
\includegraphics[width=\textwidth]{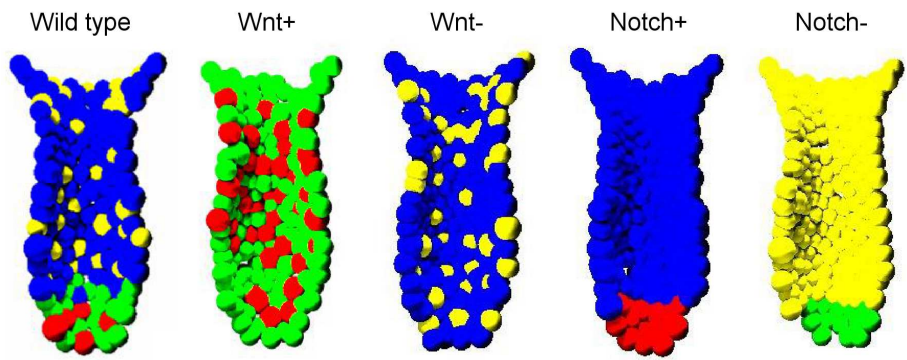}
\vspace*{8pt}
\caption{
Simulation results for the model of cell fate determination in the crypt by Buske et al.\protect\cite{Buske2011Comprehensive}, showing the spatial distribution of undifferentiated/stem (red), Paneth (green), Goblet (yellow) and enterocyte progenitor cells (blue).
Alongside the wild type case, four gain or loss of function perturbations are shown, each of which agree qualitatively with experimental observations. 
From left to right, these are: wild type; constitutive activation of Wnt in all cells (Wnt+), which results in loss of Goblet and enterocyte progenitors\protect\cite{Sansom2004Loss}; reduced Wnt signalling (Wnt-), which results in loss of undifferentiated and Paneth cells\protect\cite{Fevr2007Wnt}; constitutive activation of Notch (Notch+), which results in loss of secretory lineages\protect\cite{Fre2005Notch}; and inhibition of Notch signalling (Notch-), which results in loss of undifferentiated and absorptive cells.\protect\cite{vanEs2005Notch}
Image adapted with permission from Buske et al.\protect\cite{Buske2011Comprehensive} under a Creative Commons License.
}
\label{fig:BuskeCryptModelFigure}
\end{figure}

Buske et al.\cite{Buske2011Comprehensive} compute cellular Notch activity by summing contributions from cell neighbours and determining cell fate from a  `look-up table' that specifies distinct outcomes for particular Wnt and
Notch levels. 
This model recapitulates experimental observations on the spatial distributions of secretory Paneth and Goblet cells in wild-type crypts \cite{Chwalinksi1989Crypt,Paulus1993Differentiation}, timescales for monoclonal conversion\cite{Li1994Crypt} and the effect of Wnt/Notch mutations on crypt organization,\cite{Fevr2007Wnt,Fre2005Notch,Sansom2004Loss,vanEs2005Notch} as illustrated in Figure~\ref{fig:BuskeCryptModelFigure}. 
Other models that have sought to address the rules underlying cell fate determination and lineage specification include the work of Pin et al.\cite{Pin2012Modelling} and Wong et al.\cite{Wong2010Computational} 
These models largely restrict their focus to simple rule sets for cell regulation, rather than extending the more detailed ordinary differential equation models of Wnt signalling\cite{LloydLewis2013Towards} to account for crosstalk between pathways.

%%%%%%%%%%%%%%%%%%%%%%%%%%%%%%%%%%%%%%%%%%%%%%%%%%%%%%%

\subsection{Crypt deformations}

We conclude this section by highlighting the more recent application of discrete and multiscale modelling to study later steps in colorectal carcinogenesis.
Following the accumulation of genetic mutations which disrupt the dynamics of a crypt, excess proliferation can cause a build-up of stress, which may eventually cause the crypt to buckle and deform. 
Drasdo and Loeffler\cite{Drasdo2001Individual} were the first to investigate this process at a cellular level by modelling a vertical cross-section through a crypt as a U-shaped chain of growing, deformable elastic spheres. 
The authors demonstrate that crypt buckling can occur when cell cycle durations are reduced or the rigidity of cells is reduced. 
More recent work has sought to address the dynamic role of the basement membrane in determining crypt geometry\cite{Dunn2012Modelling,Dunn2012TwoDimensional} and the biomechanical behaviour of \textit{in vitro} crypt organoid cultures, which reproduce the coarse three-dimensional structure of crypts and allow for more precise experimental measurement, imaging and perturbation.\cite{Buske2012On}

%%%%%%%%%%%%%%%%%%%%%%%%%%%%%%%%%%%%%%%%%%%%%%%%%%%%%%%

\section{Mathematical challenges} \label{sec:challenges}

Having summarised recent efforts to apply multiscale modelling techniques to study colonic crypt organization and colorectal carcinogenesis, we next review some of the key mathematical challenges in this area.

%%%%%%%%%%%%%%%%%%%%%%%%%%%%%%%%%%%%%%%%%%%%%%%%%%%%%%%

\subsection{Inference}

An important limitation associated with multiscale models of tissue dynamics is the lack of established statistical and computational techniques with which they may be parameterised or validated against experimental data. 
It may be argued that to date, the majority of multiscale models of carcinogenesis have been used to obtain \textit{qualitative} insights into the spread and fixation of mutations within epithelial tissues. 
However, some recent studies have attempted to place such models on a more quantitative footing. 
Further, a variety of inverse problem frameworks have been developed for the estimation of model parameters. 
Key issues to be addressed in this area include how to perform model selection,\cite{Piou2009Proposing} develop simplified `emulators' to approximate such models for parameter inference,\cite{Hooten2011Assessing} and efficiently compute statistical likelihoods or perform approximate Bayesian computation (ABC).\cite{Wood2010Statistical}

While these issues have been actively pursued in the field of mathematical ecology, to date relatively few studies have considered them in the context of tissue homeostasis and carcinogenesis. 
Of particular relevance is the work of Sottriva and Tavar\'{e},\cite{Sottoriva2010Integrating} who demonstrate how ABC may be applied using methylation pattern data\cite{Yatabe2001Investigating} to infer posterior distributions for parameters of stem cell dynamics in a cellular Potts model of a colonic crypt. 
This model shares many features with the models discussed in Section~\ref{sec:multiscale_models}. 

%%%%%%%%%%%%%%%%%%%%%%%%%%%%%%%%%%%%%%%%%%%%%%%%%%%%%%%

\subsection{Computational considerations}

A further barrier to the wider use of multiscale models for the study of tissue homeostasis and carcinogenesis is the lack of standards or benchmarks. 
Despite their establishment in the cancer modelling literature, it is still the case that previously developed models and methods are seldom re-used effectively, because they are typically not available as rigorously tested, open-source simulation software. 
It is therefore difficult to guarantee the reproducibility of computational results, which is of particular importance given the complexity of such models. 

This problem is beginning to be addressed, for example by the Chaste project, which provides an open-source C++ implementation of several classes of individual cell-based and multiscale models of multicellular populations.\cite{Mirams2013Chaste} 
Chaste constitutes a multi-purpose software library that is developed to support computational simulations for a wide range of biological problems. 
It is developed using an agile approach and comprises fully tested, industrial-grade software.\cite{Mirams2013Chaste} 
A major component of this software library is the code used to simulate several of the multiscale crypt models discussed in Section~\ref{sec:multiscale_models}. 
Full details of all technical aspects of the implementation of such models, and the code, are freely available to the community (\texttt{http://www.cs.ox.ac.uk/chaste}) along with documentation and tutorials.
However, a more systematic and comprehensive description of specific multiscale models in the literature remains an ongoing challenge.

%%%%%%%%%%%%%%%%%%%%%%%%%%%%%%%%%%%%%%%%%%%%%%%%%%%%%%%

\subsection{Effect of modelling framework} \label{sec:comparing_models}

To date, there has been little comparative study of different cell-based modelling approaches. 
There are clear strengths and weaknesses associated with each approach, and therefore some situations in which it is clear which approach is most valid; for example, cellular automata are not well suited to the detailed study of cell adhesion and motility. 
However, there are other cases where it is not clear which modelling approach is valid. 
This raises the question of to what extent are the results of model simulations artifacts of the chosen modelling approach or method of numerical solution? 
It can be difficult to make an accurate comparison between different models, in order to evaluate the impact of various constitutive assumptions on the behaviour of a model. 
This difficulty is particularly relevant in the context of multiscale models, which may couple the biomechanical behaviour of cells in tissues with other processes such as the secretion, transport and uptake of nutrients or signalling molecules. 
When comparing different constitutive assumptions using the same overall modelling approach, a computational framework is required that allows one to easily change the fine details of a model and its implementation. 
A small number of studies have begun to address these issues in the context of crypt dynamics and epithelial tissue dynamics more generally.\cite{Fletcher2013Implementing,Osborne2010Hybrid}
Other work has also sought to characterise the mechanical behaviour of distinct classes of individual cell-based models under loading, unloading and shearing.\cite{Davit2013Validity,Pathmanathan2009Computational} 
Such studies could help refine the region of parameter space that must be searched when fitting such models to data.

%%%%%%%%%%%%%%%%%%%%%%%%%%%%%%%%%%%%%%%%%%%%%%%%%%%%%%%

\subsection{Coarse-graining approaches}

As the above examples demonstrate, multiscale modelling has played an increasingly important role in the study of colonic crypt organization and colorectal carcinogenesis. 
A key strength of this approach is that it allows the simulation of observed cell-level behaviours, such as cell heterogeneity and the occurrence of mutated cells. 
However, the computational cost associated with analysing such models remains a significant challenge. 
This is exacerbated by the stochastic behaviour of cell processes such as proliferation and differentiation, as well as by the increasingly detailed mathematical descriptions of cell decision-making processes. 
For example, in order to build up statistical distributions for how the probability and timescale of mutant domination varied with properties such as the location of the initial mutation in a colonic crypt, Mirams et al.\cite{Mirams2012Theoretical} needed to run a large number of realisations of their model, cumulatively equivalent to over 9000 years of simulated `crypt time'.

One approach to tackling this problem is to develop coarse-grained models that capture the essential features of the original models but are amenable to efficient simulation methods or established mathematical techniques (e.g. asymptotic or bifurcation analysis). 
As the methodology used to derive coarse-grained equations is strongly dependent on the class of underlying cell-based model, here we highlight a number of  cases that are particularly relevant to the type of crowded cellular environment found along the crypt axis. 
A common feature shared by the coarse-grained models considered below is that, in one spatial dimension, the equation describing the spatiotemporal evolution of the coarse-grained cell number density $q(y,t)$ takes the general form
\begin{align}
\frac{\partial q}{\partial t} & =\frac{\partial }{\partial y} \left(  D(q) \frac{\partial q}{\partial y}\right).
\label{NonlinearDiff}
\end{align}
Here $y$ represents position along a spatial axis and $D(q)$ is a nonlinear diffusion coefficient within which the details from the respective cell-based models appear.
%The process of coarse-graining through systematic and rational model reduction offers a crucial tool to avoid intractability.

%As discussed in Section~\ref{sec:cellular_automata}, cellular automaton models are typically defined using a set of \textit{ad hoc} rules designed to encapsulate biological understanding of aspects of system in question. Usually, coarse-graining techniques are applied for some particular constrained set of rules. For example, the lattice gas cellular automaton uses a well-defined description of cell movement and reactions that allows the application of the machinery of the lattice Boltzman equation to study the statistics of simulation behaviour. Hence, one can derive mean-field partial differential equations that allow qualitative insight into the behaviour in a simulation.\cite{Chopard2010Lattice,Hatzikirou2010Lattice}

We start by considering work by Simpson and co-workers, who have accounted for volume exclusion effects in a study of stochastic cell movement on a lattice. \cite{Baker2010Correcting,Johnston2012Mean,Simpson2007Simulating} 
By way of example, we highlight the case of stochastic excluded motion on a one-dimensional lattice. 
In this case, the authors find that the governing coarse-grained equation takes the form~\eqref{NonlinearDiff}, with the nonlinear diffusion coefficient  given by
\begin{align}
D(q) & = D_0(1-\sigma q(4-3q)),
\end{align}
where $\sigma$ is an adhesion parameter and $D_0$ represents the diffusivity of an isolated cell.

In other work, Alber and coworkers\cite{Alber2006Multiscale,Alber2007Continuous} show that the coarse-grained density derived from an underlying lattice-based cellular Potts model is also described by~\eqref{NonlinearDiff}, with the nonlinear diffusion coefficient given by
\begin{align}
D(q)=\frac{D_0}{(K-q)^2},
\end{align}
where $K$ is a parameter that defines the maximum allowed cell density on the lattice and $D_0$ is again the diffusion coefficient representing the random motion an isolated cell. 
Interestingly, as the density tends to the limiting value $K$, the diffusion coefficient tending to infinity represents the case of maximal cell packing.

Considering off-lattice cell-based models, Murray and co-workers\cite{Murray2009From,Murray2012Classifying} show how different inter-cellular force laws (e.g. linear, Hertz, Lennard-Jones) can be described using a range of corresponding nonlinear diffusion equations. As before, in one spatial dimension the governing equation takes the form \eqref{NonlinearDiff} and in the case of a linear spring force between cells, the diffusion coefficient takes the form
\begin{align}
D(q) & = \frac{k }{\eta q^2},
\end{align}
where $k$ and $\eta$ denote the spring constant and damping coefficient, respectively. 
Subsequent studies have performed a similar derivation starting from alternative stochastic off-lattice cell-based models.\cite{Fozard2010Continuum,Middleton2014Continuum}
 
\begin{figure}
\centering
\includegraphics[width=0.8\textwidth]{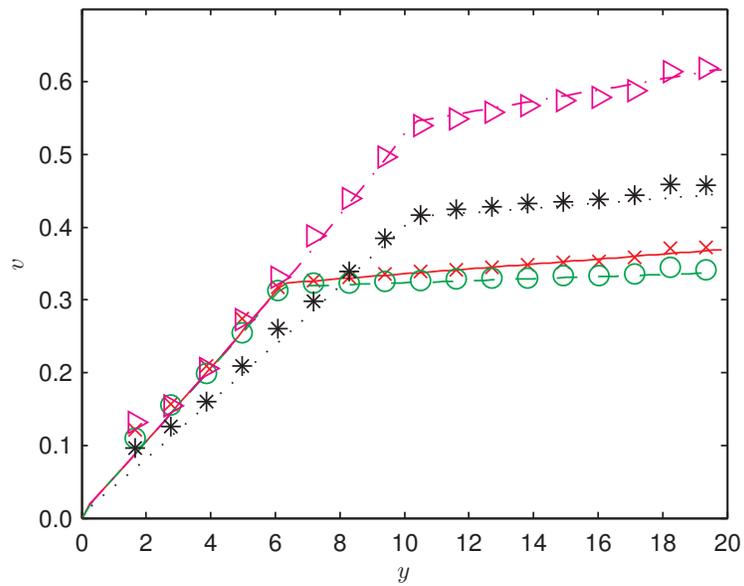}
\vspace*{8pt}
\caption{Steady-state velocity plotted as a function of crypt axis for different values of the spring constant, $k$, and cell cycle period $T_C$. Image reproduced with permission from Murray et al.\protect\cite{Murray2012Classifying}. Markers  denote measurements from simulations of cell-based model. Lines denote solutions from continuum models.}
\label{fig:MurrayCryptCtm}
\end{figure}

The work of Murray and coworkers\cite{Murray2009From,Murray2010Modelling,Murray2011Comparing} has been applied directly to the intestinal crypt models proposed by Meineke et al.\cite{Meineke2001Cell} and van Leeuwen et al.\cite{vanLeeuwen2009Integrative} as described in Section~\ref{sec:off_lattice_models}. 
We therefore explore their model in further detail. 
The authors focus on the simulation framework described in Section~\ref{sec:off_lattice_models}: a crypt of length $L$ is populated by cells of equilibrium diameter $a$ and treated as an unwrapped cylinder. 
Exploiting the axisymmetric nature of the crypt, the authors consider a model in one spatial dimension and show that the cell number density along the crypt axis, $q(y,t)$, satisfies the nonlinear diffusion equation
\begin{align}
\label{eq:philip} \frac{\partial q}{\partial t} & =\frac{\partial }{\partial y}\left(  D(q) \frac{\partial q}{\partial y} \right) + \frac{\ln 2}{T_C}q H(y_{crit}-y), 
\end{align}
where $D(q)=k/\eta q^2$,  $T_C$ is the mean cell cycle period in the proliferative region of the crypt, $y_{crit}$ is the height up the crypt below which cells can proliferate and $H(.)$ is the Heaviside function. 
In Figure~\ref{fig:MurrayCryptCtm} we compare measurements of the average cell velocity profile along the crypt axis from the underlying cell-based model with predictions from the continuum model for different points in parameter space. Despite the simplifying assumptions made in the derivation of equation \eqref{eq:philip}, it can be seen that the coarse-grained model captures qualitatively the behaviour of the multi scale model.

Upon deriving equation~\eqref{eq:philip}, it is natural to apply standard techniques of analysis. 
For example, a critical non-dimensional parameter is identified by the authors; seeking a steady-state solution in which the rate of proliferation in the base of the crypt is matched by the rate of loss at the top, the authors identify that the condition
\begin{align}
k > \frac{\eta \ln 2 y_{crit}(2L-y_{crit})}{a^2 T_C}
\end{align}
must hold in order that a steady-state solution exists. 
Whilst this result was verified using simulations of the underlying cell-based model, identifying such a parameter relationship using numerical simulations alone would be a non-trivial computational task. It is notable from the above examples that the nonlinear diffusion equations that have arisen from a number of disparate coarse-graining approaches provide a unifying framework with which to classify and characterise behaviour in the underlying cell-based model. 
A major mathematical challenge in this endeavour is to extend such work to more detailed two- and three-dimensional tissue geometries.

%%%%%%%%%%%%%%%%%%%%%%%%%%%%%%%%%%%%%%%%%%%%%%%%%%%%%%%

\section{Concluding remarks} \label{sec:conclusions}

This review has highlighted the role that multiscale mathematical modelling can play in gaining mechanistic insights into the spatiotemporal organization of intestinal crypts under normal and pathological conditions. 
We have highlighted major mathematical and computational challenges associated with the multiscale modelling approach, concluding with a discussion of recent efforts to obtain coarse-grained descriptions of such models. 
Such attempts offer one possible way of leveraging the well-developed tools of mathematical analysis for continuous models to address key problems in multiscale modelling. 

%%%%%%%%%%%%%%%%%%%%%%%%%%%%%%%%%%%%%%%%%%%%%%%%%%%%%%%

%\section*{Note Added}
%Should be placed before Acknowledgment.

%%%%%%%%%%%%%%%%%%%%%%%%%%%%%%%%%%%%%%%%%%%%%%%%%%%%%%%

\section*{Acknowledgment}
A.G.F. is supported by EPRSC through grant EP/I017909/1.

%%%%%%%%%%%%%%%%%%%%%%%%%%%%%%%%%%%%%%%%%%%%%%%%%%%%%%%

%%%%%%%%%%%%%%%%%%%%%%%%%%%%%%%%%%%%%%%%%%%%%%%%%%%%%%%

\end{document}